\documentclass[twocolumn,showpacs,preprintnumbers,amsmath,amssymb,prl]{revtex4}

\usepackage{graphicx}% Include figure files
\usepackage{dcolumn}% Align table columns on decimal point
\usepackage{bm}% bold math

\newcommand{\be}{\begin{equation}}
\newcommand{\ee}{\end{equation}}
\newcommand{\bea}{\begin{eqnarray}}
\newcommand{\eea}{\end{eqnarray}}
\newcommand{\df}{{\rm d}}

\begin{document}

\preprint{Superluminal group velocity \ldots}

\title{Superluminal group velocity of neutrinos}

\author{Antonio Mecozzi}%
\email{antonio.mecozzi@univaq.it}
\affiliation{%
Department of Electrical and Information Engineering, University of L'Aquila, L'Aquila, Italy}

\author{Marco Bellini}%
\email{bellini@ino.it}
\affiliation{%
Istituto Nazionale di Ottica Applicata (INO-CNR), L.go E. Fermi 6, 50125 Florence, Italy}

\date{\today}

\begin{abstract}

We suggest a possible interpretation of the recent observation by the OPERA collaboration of superluminal propagation of neutrinos. We show that it is in principle possible that the group velocity of neutrinos exceeds the speed of light without violating special relativity.

\end{abstract}

%\keywords{}
\maketitle

\section{Introduction}

Recent results of the OPERA collaboration \cite{Opera} have suggested that the speed of neutrino may exceed the speed of light in vacuum. Although this may at first sight appear inconsistent with special relativity, there are cases where superluminal propagation has been predicted and indeed observed in the past, and these prediction and observation did not violate special relativity. This is because special relativity does not prevent that a waveform can propagate with a group velocity exceeding the speed of light in vacuum, but only that a signal cannot be transmitted faster than light. To experience superluminal propagation, a waveform should experience large dispersion and intra-waveform interference. In optics, this dispersion may be provided by an interferometer, like a Mach-Zehnder interferometer, or by other mechanisms. The constructive and destructive interference taking place at the wavefront and at the trailing edge of the pulse is responsible for the superluminal propagation. With neutrinos, the large dispersion may be provided by the neutrino oscillations between different flavors,  producing constructive and destructive interferences between the different neutrino energies in which a neutrino wave-packet can be decomposed.

\section{Neutrino propagation}

The Hamiltonian for the coupling of the two neutrino flavors in vacuum in the ultra-relativistic limit is \cite{Metha,Garcia}
\be \mathbf{H} = \left( pc + \frac{\epsilon_0} 2 \right) \mathbf 1 + \frac \epsilon 2 [\sin (2 \Theta) \sigma_x - \cos(2 \Theta) \sigma_z  ] , \ee
where 
\be  \epsilon_0 = \frac{(m_1^2 + m_2^2) c^4}{2 p c}, \quad \epsilon =  \frac{(m_2^2 - m_1^2) c^4}{2 pc}, \ee
$p \simeq E/c$ is the momentum ($E$ the energy) of the neutrino, $\Theta$ is the mixing angle in vacuum. The eigenstates of the energy are
\be E_{\pm} = \left( pc + \frac{\epsilon_0}{2} \right) \pm \frac \epsilon 2 = E_0 \pm \frac{\epsilon} 2, \ee
and the eigenstates
\bea |p, \phi_-\rangle &=& -\cos(\Theta) |p, \phi_1\rangle + \sin(\Theta) |p, \phi_2\rangle, \\
|p, \phi_+\rangle &=& \sin(\Theta) |p, \phi_1\rangle + \cos(\Theta) |p, \phi_2\rangle, \eea
where $| p, \phi_i \rangle = |p\rangle \otimes |\phi_i\rangle$ are simultaneous eigenstates of momentum and flavor.

The $|\psi_0\rangle$ at time $t =0$ evolves at time $t$ into the state $|\psi_t\rangle$ given by
\bea |\psi_t\rangle &=& \int \df p' \exp(-iE_0 t/\hbar) \nonumber \\
&& \big[\langle p', \phi_+|\psi_0\rangle \exp\left(-i \epsilon t/2 \hbar \right) |p', \phi_+\rangle \nonumber \\
&& + \langle p', \phi_-| \psi_0\rangle \exp\left(i \epsilon t /2 \hbar\right) |p', \phi_-\rangle \big]. \eea
There is a one-to-one analogy to the response of a Mach-Zehnder interferometer, and to other interference effects like those taking place in fibers with polarization mode dispersion. All these effects are long-time known to show superluminal propagation \cite{Wang,Moreno,Gisin}.

Assume that we start with one neutrino flavor, that is with $\langle p, \phi_2 |\psi_0\rangle = 0$, and that the initial amplitude of the neutrino waveform is $\langle p, \phi_1 |\psi_0\rangle = \langle p |\psi_0\rangle$. We have
\bea |\psi_t\rangle &=& \int \df p' \, \langle p' |\psi_0\rangle \exp(-iE_0 t/\hbar) \nonumber \\
&& \big[ \sin(\Theta) \exp\left(-i \epsilon t/2 \hbar \right) |p', \phi_+\rangle \nonumber \\
&& - \cos(\Theta) \exp\left(i \epsilon t /2 \hbar\right) |p', \phi_-\rangle \big]. \label{70} \eea
\begin{figure}[ht]
\includegraphics[width=1 \columnwidth]{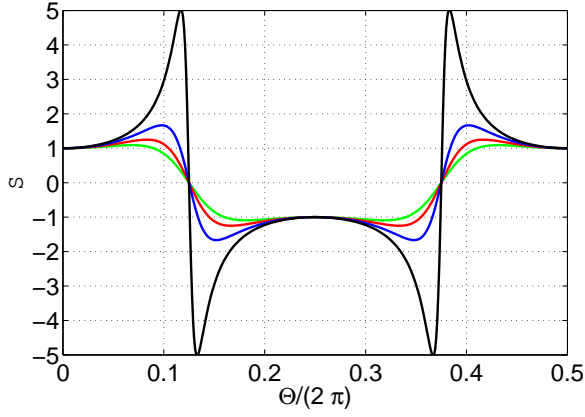}
\caption{Factor $\mathcal S$ vs. $\Theta$, for $\sin^2\left(\epsilon t/2 \hbar\right) = 0.99$ (black curve), 0.9 (blue curve), 0.8 (red curve) and 0.7 (green curve)}
\label{Fig1}
\end{figure}
Assume also that we measure the position of the neutrinos and their flavor, and that our detector is sensitive to neutrinos with flavor 1 so that we may average the position of the neutrino only when they appear in flavor 1. The measurement is equivalent to a flavor measurement followed by a position measurement on the collapsed state after the successful detection of flavor 1. The collapsed state when flavor 1 is detected is
\be |\psi_t'\rangle = \frac 1 {\sqrt{\mathcal D}} \int \df p' \, | p', \phi_1 \rangle  \langle p', \phi_1 |\psi_t\rangle, \ee
where the normalization factor is
\be \mathcal D = \int \df p' \langle \psi_t |p', \phi_1\rangle \langle p', \phi_1| \psi_t \rangle. \ee
The position measurement on the collapsed state gives the average
\be \langle x_t\rangle = \int \df p' \, \langle \psi'_t| p' \rangle \left(i \hbar \frac{\partial}{\partial p'} \right) \langle p' |\psi_t'\rangle = \frac {\mathcal N} {\mathcal D}, \ee
where
\be \mathcal N = \int \df p' \langle \psi_t| p', \phi_1 \rangle  \left(i \hbar \frac{\partial}{\partial p'} \right) \langle p', \phi_1 |\psi_t\rangle. \label{110} \ee
Using Eq. (\ref{70}) we obtain 
\bea \langle p', \phi_1|\psi_t\rangle = \langle p' |\psi_0\rangle F(p'), \eea
where
\bea F(p') &=& \exp(-iE_0 t/\hbar)  \big[\sin^2(\Theta) \exp\left(-i  \epsilon t/2 \hbar \right) \nonumber \\
&& + \cos^2(\Theta) \exp\left(i \epsilon t /2 \hbar\right) \big]. \eea
Assume now that the initial distribution of the neutrino momentum $|\langle p' |\psi_0\rangle|^2$ is narrow with respect to $F(p')$, and centered on $p' = p$. Then we have
\be \mathcal D = \int \df p' \langle \psi_t |p', \phi_1\rangle \langle p', \phi_1| \psi_t \rangle \simeq |F(p)|^2. \ee
Let us now set in Eq. (\ref{110}) $\mathcal N = \mathcal N_1 + \mathcal N_2$, where
\be \mathcal N_1 = \int \df p' |F(p')|^2  \langle \psi_0|p'\rangle \left(i \hbar \frac{\partial}{\partial p'} \right) \langle p' |\psi_0\rangle, \label{150} \ee
and
\be \mathcal N_2 = \int \df p' |\langle p' |\psi_0\rangle|^2 F(p')^* \left(i \hbar \frac{\partial}{\partial p'} \right) F(p). \label{160}  \ee
Integration by parts of the integral at right hand side of Eq. (\ref{150}) gives for the imaginary part of $\mathcal N_1$
\be \mathrm{Im} (\mathcal N_1) = - \frac \hbar 2 \int \df p' |\langle p' |\psi_0\rangle|^2 \frac{\partial}{\partial p'} |F(p')|^2. \ee
Using now that $|\langle p' |\psi_0\rangle|^2$ is narrow-band around $p' = p$, we obtain
\be \mathcal N_1 \simeq |F(p)|^2 \langle x_0 \rangle - \frac {i \hbar} 2 \frac{\partial}{\partial p} |F(p)|^2, \ee
where
\be \langle x_0\rangle = \int \df p \langle \psi_0| p \rangle  \left(i \hbar \frac{\partial}{\partial p} \right) \langle p|\psi_0\rangle. \ee
Using once again the narrow-band condition on the right hand side of Eq. (\ref{160}), we obtain for $\mathcal N_2$ 
\be \mathcal N_2 \simeq F(p)^* \left(i \hbar \frac{\partial}{\partial p'} \right) F(p). \ee
Performing the differentiations, we finally obtain
\be \langle x_t\rangle = \langle x_0\rangle + v_g t, \ee
where the group velocity is
\be v_g = \frac{\partial E_0}{\partial p} + \frac{1}{2 \mathcal D}  \left[\sin^4(\Theta)  - \cos^4(\Theta) \right] \frac{\partial \epsilon}{\partial p}, \ee
and
\be \frac{\partial \epsilon}{\partial p} = - \frac{\epsilon}{p}. \ee
Straightforward algebraic manipulations give
\be v_g = \frac{\partial E_0}{\partial p} + \mathcal S\, \frac{\epsilon}{2 p} = c - \frac{\epsilon_0}{2 p} + \mathcal S\, \frac{\epsilon}{2 p}, \ee
where we defined
\be \mathcal S = \frac{\cos(2 \Theta)}{1 - \sin^2(2 \Theta) \sin^2\left(\epsilon t/2 \hbar\right)}. \ee
Assume that the propagation time $t$ is such that $\sin\left(\epsilon t/2 \hbar\right) \simeq 1$. For $\sin^2\left(\epsilon t/2 \hbar\right) > 0.5$, there are regions of $\Theta$ where $\mathcal S$ is larger in modulus than one (see Fig. \ref{Fig1}). Assume now that one of the two masses is much larger than the other. If $m_2 \gg m_1$, then $\epsilon \simeq \epsilon_0 >0$ and in the regions where $\mathcal S >1$ we have $v_g > c$. If instead $m_2 \ll m_1$, then $\epsilon \simeq - \epsilon_0 < 0$ and $v_g > c$ where $\mathcal S <-1$. In both cases, we obtain superluminal propagation. Of course, this does not mean that the speed of a possible signal transmitted with a neutrino wave-packet exceeds the speed of light, it is just a property that comes from the wave-packet deformation caused by the interference of the two possible quantum paths that a neutrino may follow before reaching the detector.

\section{Conclusions}

We have shown that a simple model of two flavor neutrino interaction predicts that there are regimes where a neutrino waveform may travel with a group velocity exceeding the speed of light. Although our model may be a simplistic one, other possibly unknown interactions may account for a group velocity of the neutrino waveform exceeding the speed of light, all involving some form of dispersive loss depressing the trailing edge of the neutrino waveform. The bottom line here is that the observation of a group velocity of a neutrino wave-packet larger than the speed of light does not necessarily contradict special relativity.

\section{Acknowledgments}

We would like to thank Prof. Kazuyuki Fujii, Department of Mathematical Sciences, Yokohama City University, for noting an algebraic error in a previous version of this Letter.

\end{document}